\documentclass[twocolumn]{openjournal} 

\usepackage{orcidlink} 
\usepackage[T1]{fontenc}
\usepackage{ae,aecompl}
\usepackage[T1]{fontenc}
\usepackage{ae,aecompl}
\usepackage{hyperref}
\usepackage{url}

\usepackage{longtable}
\usepackage{booktabs}
\usepackage{tabu}
\usepackage{graphicx}
\usepackage{multirow}
\usepackage{ulem}
\usepackage{color}
\usepackage{amsmath}

\def \rmModot{~\rm{M_\odot}}

\def \cm{~\rm{cm}}
\def \s{~\rm{s}}
\def \km{~\rm{km}}

\def \g{~\rm{g}}

\def \erg{~\rm{erg}}

\usepackage{xcolor}
\definecolor{redak}{rgb}{0.9,0.15,0.05}

\shorttitle{Double shells in SN 2024ggi}
\shortauthors{Shiran and Soker}

\graphicspath{{./}{figures/}}

\begin{document}

\title{Double shell structure in supernova 2024ggi}

\author{Kobi Shiran\,\orcidlink{0009-0008-3236-1210}}
\affiliation{Department of Physics, Technion Israel Institute of Technology, Haifa, 3200003, Israel; kobishiran@campus.technion.ac.il}

\author{Noam Soker\,\orcidlink{0000-0003-0375-8987}} 
\affiliation{Department of Physics, Technion Israel Institute of Technology, Haifa, 3200003, Israel; soker@technion.ac.il}


\begin{abstract}
We built a simple toy model of a core-collapse supernova (CCSN)  ejecta composed of two shells, an outer low-mass spherical shell and an inner elongated massive shell, and show that it can reproduce the evolution of the photospheric radius of SN 2024ggi, $R_{\rm ph}(t)$. During the first week, the larger spherical shell, the S-shell, forms the photosphere. As the shell expands and becomes increasingly transparent, the photosphere moves inward along the mass coordinate, although it grows in size. When the photosphere reaches the long axis of the elongated inner shell, the E-shell begins to contribute to the photosphere, ultimately comprising the entire photosphere. The simple toy model explains the transition of $R_{\rm ph}(t)$ from being concave (decreasing slope) to convex (increasing slope). A single-shell model predicts only concave behavior.  The structure of a spherical shell with an inner elongated shell is motivated by the morphologies of several CCSN remnants whose structures have been attributed to multiple pairs of jets in the framework of the jittering jets explosion mechanism (JJEM).   The deduced multiple-shell ejecta of SN 2024ggi in this study, and of SN 2023ixf in an earlier study, as well as studies of the polarization of SN 2024ggi, are better compatible with the JJEM than with the neutrino-driven mechanism. Our study supports the growing evidence that the JJEM is the primary explosion mechanism of CCSNe.     
\end{abstract}




\section{Introduction} 
\label{sec:intro}

There are two intensively studied competing theoretical explosion mechanisms of core-collapse supernovae (CCSNe) aiming at explaining most, or even all, CCSNe: the jittering-jets explosion mechanism (JJEM; \citealt{Soker2024UnivReview, Soker2025Learning} for recent reviews\footnote{See \citealt{Soker2025Padova} for a talk on the JJEM: \url{https://www.memsait.it/videomemorie/volume-2-2025/VIDEOMEM_2_2025.47.mp4}}) and the delayed neutrino-heating (neutrino-driven) mechanism (\citealt{Janka2025} for a recent review\footnote{See  \citealt{Janka2025Padova} for a recent talk on the neutrino-driven mechanism: \url{https://www.memsait.it/videomemorie/volume-2-2025/VIDEOMEM_2_2025.46.mp4}}). 
Other energy sources, like a magnetar and fallback accretion, operate after the explosion and might provide additional energy to the exploding massive star (see \citealt{Soker2026G11} on the relation of these to the two explosion mechanisms). 

The observable property that decisively distinguishes between the two explosion models is the point-symmetric type morphologies of CCSN remnants (CCSNRs). The JJEM predicts that many, but not all, CCSNRs possess point-symmetric morphologies that are shaped by two or more pairs of opposite jets that do not share the same symmetry axes. These pairs of jets exploded the star; neutrino heating boosts the explosion, but it is not the primary energy source (\citealt{Soker2022nu}).   
Therefore, the research of the JJEM has focused since 2024 on finding supporting evidence for jets in CCSNRs (e.g., papers since 2025, \citealt{Bearetal2025Puppis, BearSoker2025, Shishkinetal2025S147, Soker2025G0901, Soker2025N132D, Soker2025RCW89, Soker2025Dust, SokerAkashi2025, SokerShishkin2025Vela, SokerShishkin2025W49B, Soker2026G11}). 
\cite{Braudoetal2025} demonstrated the shaping of point-symmetric morphologies with three-dimensional hydrodynamical simulations in the framework of the JJEM, and \citep{WangShishkinSoker2025} showed that striped-envelope CCSNe also have pre-collapse convective perturbations that seed the formation of stochastic angular momentum accretion, which in turn leads to the launching of pairs of jittering jets.  

On the other hand, the neutrino-driven mechanism cannot explain all aspects of point-symmetric CCSNRs \citep{SokerShishkin2025Vela}. Therefore, point-symmetric CCSNRs strongly suggest that the JJEM is the primary explosion mechanism for CCSNe. 
A point symmetric circumstellar matter or post-explosion jets, both of which might in principle operate with the neutrino-driven explosion mechanism, can at best explain a minority of structural pairs in some CCSNRs, but fail to explain most point-symmetric CCSNR morphologies (e.g., \citealt{SokerShishkin2025Vela}). The neutrino-driven mechanism studies mainly simulate the revival of the stalled shock at $\simeq 150 \km$ from the newly born neutron star (NS) with neutrino heating, find the conditions for explosions, and compare simulations with some other observations (e.g., \citealt{Bambaetal2025CasA, Bocciolietal2025, BoccioliRoberti2025, EggenbergerAndersenetal2025, FangQetal2025, Huangetal2025, Imashevaetal2025, Laplaceetal2025, Maltsevetal2025, Maunderetal2025, Morietal2025, Mulleretal2025, Nakamuraetal2025, SykesMuller2025, Orlandoetal20251987A, ParadisoCoughlin2025, PowellMuller2025, Tsunaetal2025, Vinketal2025, WangBurrows2025, Willcoxetal2025, Mukazhanov2025, Raffeltetal2025, Vartanyanetal2025, Calvertetal2025, LuoZhaKajino2026}). The magnetorotational explosion mechanism requires a rapidly rotating pre-collapse core. It therefore operates in very rare cases by launching one pair of jets along a fixed axis (e.g., \citealt{Shibataetal2025}). It might account for only a small fraction of CCSNe. Studies of the magnetorotational explosion mechanism attribute most CCSNe to the neutrino-driven mechanism, so we group the magnetorotational mechanism with it.  

Observations during the explosion process of CCSNe and during the photospheric phase, which occurs at the first several weeks to a few months after explosion, provide only a few observables to distinguish between the two explosion mechanisms, as the two have similar predictions to most observables during this phase, i.e., before the ejecta is spatially resolved (e.g., for reviews \citealt{Soker2024UnivReview, Soker2025Learning}).
One possible property is the explosion energy. The small number of CCSNe with explosion energies of $E_{\rm ex} \gtrsim 2 \times 10^{51} \erg$ (as some superluminous stripped-envelope supernovae; e.g., \citealt{Kumar2025}) supports the JJEM because the neutrino-driven mechanism struggles to reach these explosion energies. 
Another emerging property during the photospheric phase is the presence of multiple photospheric shells. 

In \cite{SokerShiran2025}, we analyzed the photosphere radius evolution $R_{\rm ph}(t)$ of SN 2023ixf that \cite{Zimmermanetal2024} calculated, and found that we can explain the evolution with a structure of two, and possibly three, photospheric shells. We noted there that the morphologies of several CCSNRs exhibit two or more complete or partial shells that may form such photospheric shells shortly after explosion. Studies have attributed shell morphologies to jet shaping within the framework of the JJEM. We concluded that a photospheric shell structure is consistent with and supports the JJEM. 
In this study, we consider a double-shell structure in which the outer shell is large-scale and spherical, whereas the inner shell is elongated. This is motivated by the structure of some CCSNRs, two of which we present in Section \ref{sec:TwoShells}. 
Based on these two-shell morphologies, in Section \ref{sec:Method} we build a simple toy model to calculate the observed photospheric radius. In Section \ref{sec:Photosphere} we fit a two-shell model to the photospheric radius evolution of SN 2024ggi. 
We summarize this short study by further strengthening the JJEM in Section \ref{sec:Summary}.

\section{Two-shells morphology of CCSNRs}
\label{sec:TwoShells}

\begin{figure}[th]
\begin{center}
\includegraphics[trim=0.1cm 13.00cm 6.8cm 2.3cm ,clip, scale=1.05]{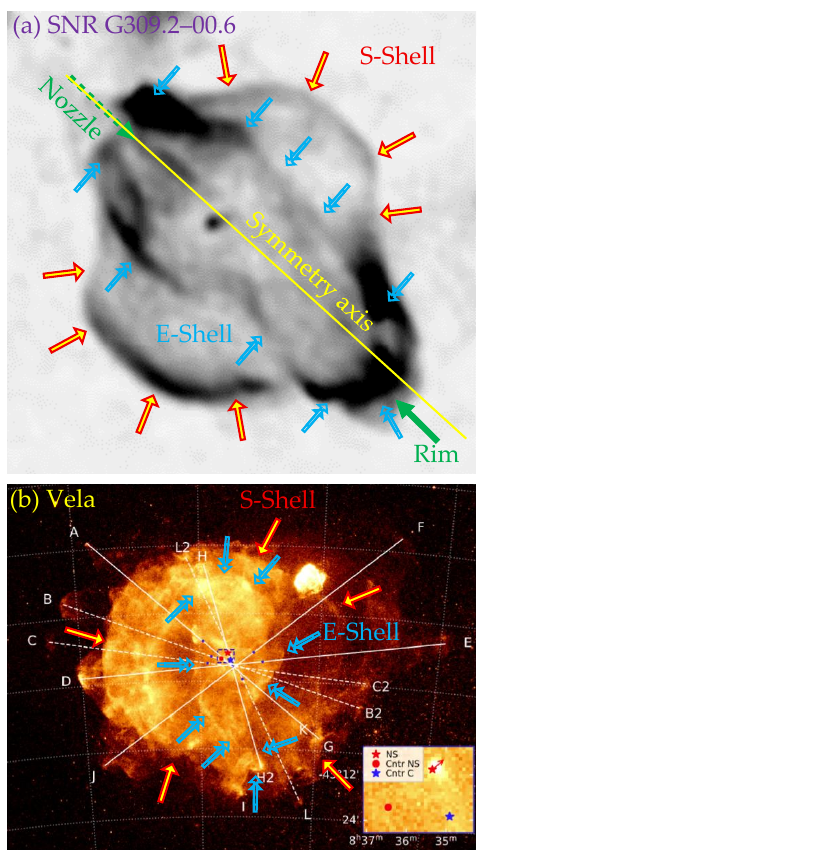}
\caption{ Two images of CCSNRs with an elongated shell inside a large-scale, more-or-less spherical one. Studies attributed their morphologies to the JJEM. In each image, we marked the two prominent shells that can form two photospheric shells: as the outer, large-scale spherical shell, S-shell (marked with the yellow-red arrows), becomes transparent, the inner, elongated one, E-Shell (marked by the double-headed pale-blue arrows), takes over. (a) A radio image of SNR G309.2–00.6 adapted from \cite{Gaensleretal1998}, who noted the jet-shaped morphology. \cite{Soker2024PNSN} identified the rim-nozzle symmetry. (b) An eROSITA DR1 (log scale, $0.2-2.3~\rm{keV}$) X-ray counts image of the Vela CCSNR adapted from \cite{SokerShishkin2025Vela}. The lines depict the point-symmetric structure of Vela: dashed lines represent pairs of opposite structural features identified by \cite{SokerShishkin2025Vela}, while the solid lines represent earlier-identified pairs. Blue dots are the centers of the lines, and the blue asterisk is the center of these dots. The inset on the bottom right ($29.2^{\prime} \times 22.7^{\prime}$) is the inner part of the Vela SNR, including the NS location (\citealt{Kochanek2022}; red asterisk), its projected movement direction (red arrow), and the presumed origin at explosion (\citealt{Kochanek2022, Dodson_etal_2003}; red dot).}
\label{fig:SN2024ggiFigureCCSNRs}
\end{center}
\end{figure}

To motivate the simple two-shells photospheric model (Section \ref{sec:Method}), we present two CCSNRs in Figure \ref{fig:SN2024ggiFigureCCSNRs}.  

Panel (a) of Figure \ref{fig:SN2024ggiFigureCCSNRs} presents a radio image that we adapted from \cite{Gaensleretal1998}, who already argued that jets shaped this CCSNR. \cite{Soker2024PNSN} identified a nozzle-rim symmetry axis and attributed the shaping and explosion to the JJEM. In line with our goals, we note two prominent shells. An outer one that crudely has a large-scale spherical structure. The red-yellow arrows point at eight locations on this shell's projection onto the plane of the sky; we term it the S-shell. We also identify an elongated shell (which we will model as an ellipsoidal shell) whose small axis is much smaller than the S-shell diameter, but whose long axis is much larger than the S-shell diameter. 

Panel (b) of Figure \ref{fig:SN2024ggiFigureCCSNRs} presents an 
eROSITA X-ray image of the Vela CCSNR that we adapted from \cite{SokerShishkin2025Vela}. This CCSNR has a rich point-symmetric morphology composed of the following prominent structures. There is a crude, large-scale spherical shell, which we mark with the yellow-red arrows; this is the S-Shell of Vela. There are pairs of clumps, as depicted by the dashed and solid lines. In addition to these, the eROSITA image reveals an `S-shaped' structure composed of heavy metals \citep{SokerShishkin2025Vela}. The outer boundary of this `S-shaped' structure is the elongated shell, the E-shell, that we mark with the double-headed pale blue arrows. Although the E-Shell in Vela does not have a large-scale ellipsoidal structure, in this study, it is adequate to model the E-Shells as ellipsoids. 

The CCSNRs W44 (attributed to the JJEM by \citealt{Soker2024W44}) and the Cygnus Loop (attributed to the JJEM by \citealt{ShishkinKayeSoker2024}) also have an inner elongated shell inside a more spherical outer one. These four CCSNRs, which were clearly shaped by energetic jets, i.e., the jets that exploded the respective CCSNe in the frame of the JJEM, motivate us in the modeling we describe next.

\section{The numerical method}
\label{sec:Method}

\cite{ChenXetal2024} calculated the photospheric radius of SN 2024ggi, $R_{\rm ph} (t)$, which we will present in Section \ref{sec:Photosphere}. 
The time evolution of $R_{\rm ph}(t)$ for SN 2024ggi exhibits two early-time phases, both of which grow almost linearly (but with a significant deviation) with time, but with markedly different slopes. A single shell cannot explain this sharp change in behavior, whether it is spherical or elongated (such as an ellipsoid). 
       
\begin{figure}[th]
\begin{center}
\includegraphics[trim=5.5cm 2.0cm 5.0cm 2.0cm ,clip, scale=0.5]{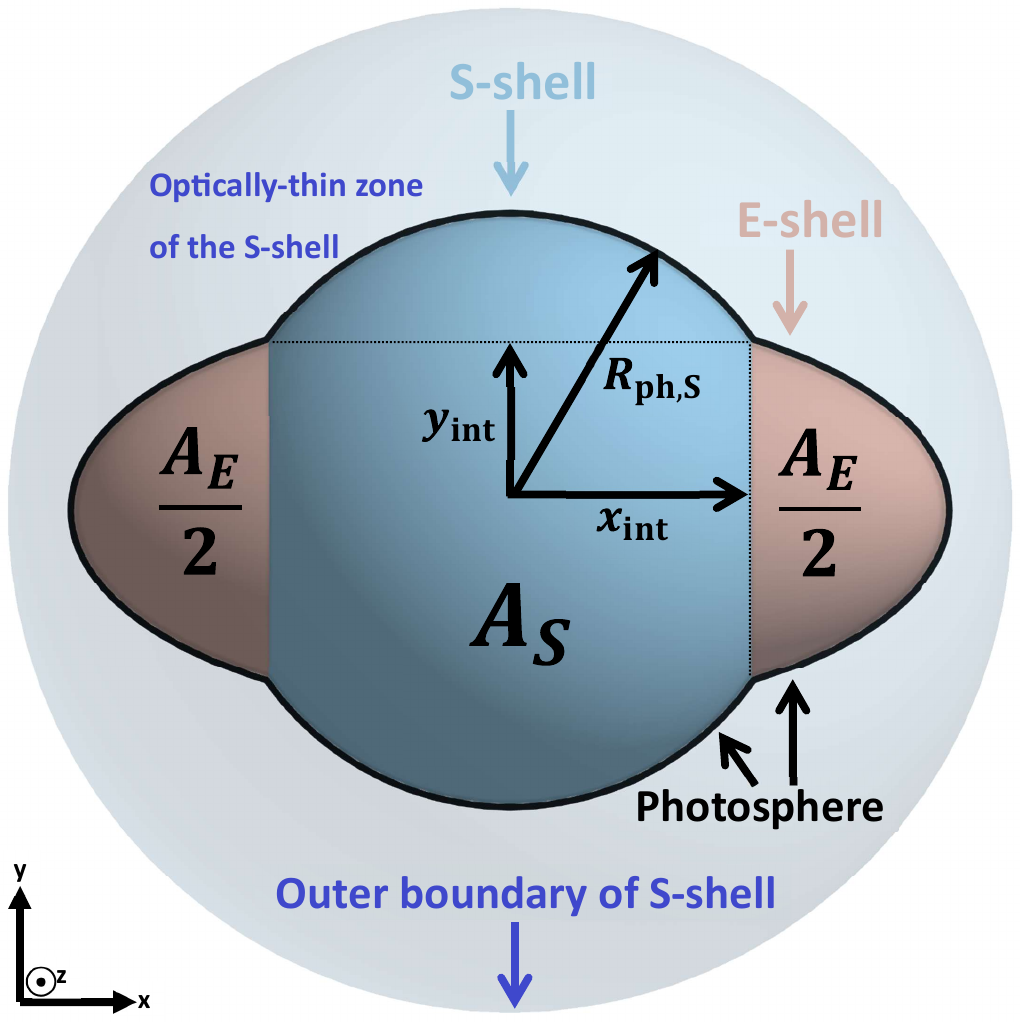}
\caption{A schematic look at the photosphere of our toy model during a time when both shells contribute to the photosphere: the blue is the S-shell with a projected (on the plane of the sky) area $A_{\rm S}$, and the brown is the E-shell with a projected area $A_{\rm E}$. The thick black line represents the photospheric limb. The line of sight is perpendicular to the long axis of the E-shell; there is an axial symmetry around this $x$-axis. The physical size of the S-shell is larger than that of the E-shell, but at this time, the outer S-shell zone is optically thin, as indicated.  }
\label{fig:SN2024ggiFigureVisualization}
\end{center}
\end{figure}

\begin{figure}[th]
\begin{center}
\includegraphics[trim=0.0cm 0.00cm 0.0cm 0.0cm ,clip, scale=0.2]{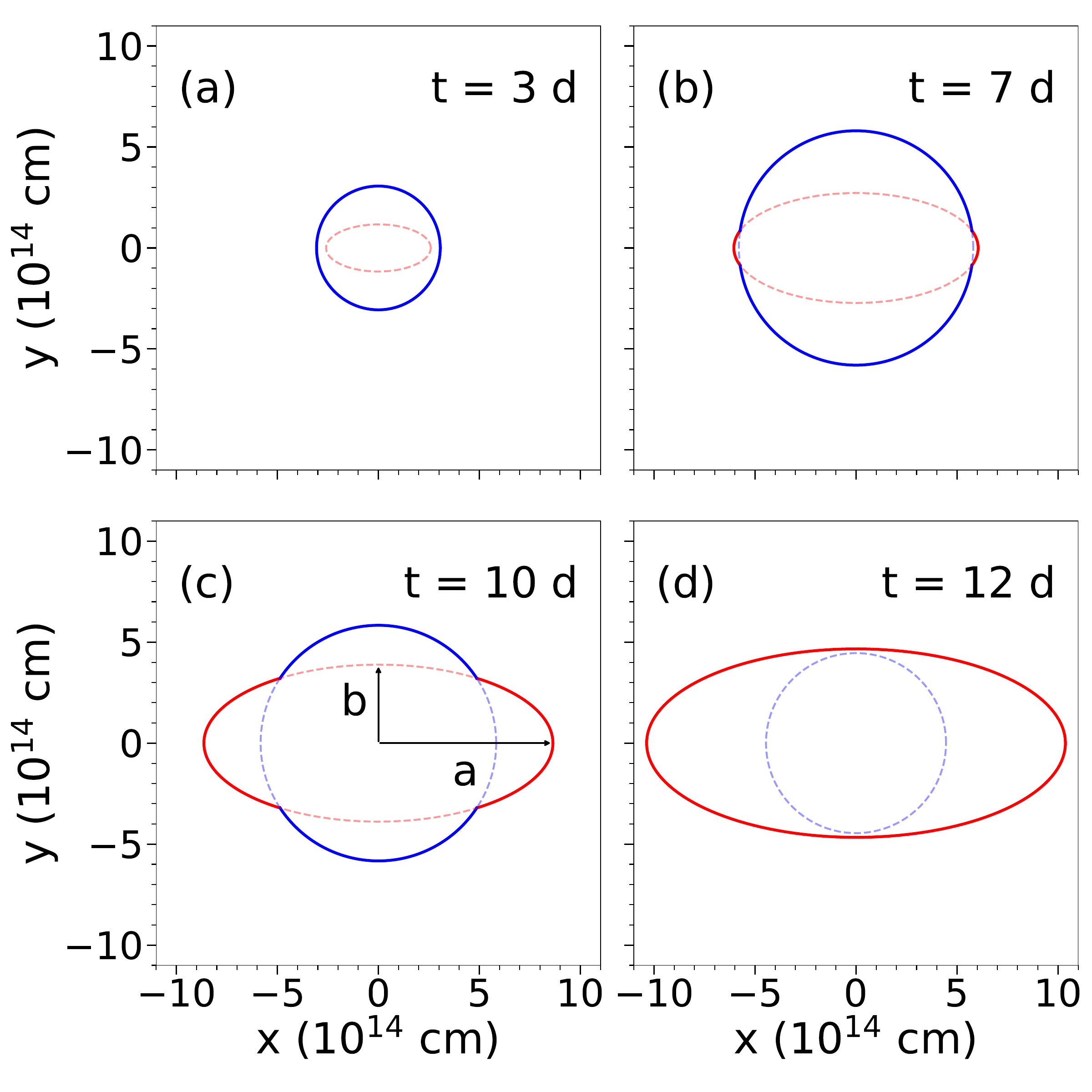}
\caption{The photosphere limb at four times depicted by the solid lines: blue for the S-shell and red for the E-shell. Dash-blue line is the rest of the S-shell photosphere had there been no E-shell; dash-red line is the rest of the E-shell photosphere had there been no S-shell. Each shell is expanding at a constant velocity (Homologous expansion). (a) Only the S-shell contributes to the photosphere, as it is optically thick. (b) The outer S-shell becomes optically thin, and the E-shell starts contributing to the photosphere. (c) The E-shell contributes a significant fraction of the photosphere. (d) The E-shell forms the entire photosphere. 
The sizes of the shells and the photosphere structure presented here at the four times are of the fiducial toy model that we further describe in Section \ref{sec:Photosphere}.  }
\label{fig:SN2024ggiFigureSnapshots}
\end{center}
\end{figure}

We construct a toy model comprising a spherical shell (S-shell) and an elongated prolate ellipsoid (E-shell), as schematically shown in Figure \ref{fig:SN2024ggiFigureVisualization}. Both expand homologously and maintain their shape. The S-shell expands faster than the E-shell and contains less mass. At early times, the photosphere is the outer boundary of the S-shell and expands linearly with time, as panel (a) of Figure \ref{fig:SN2024ggiFigureSnapshots} schematically presents. Within a few days (panel b of Figure \ref{fig:SN2024ggiFigureSnapshots}), its outer density drops sufficiently that the photosphere moves inward in mass coordinate, eventually receding to meet the outer boundary of the E-shell. The E-shell starts to surpass the S-shell and `takes over', so that the fraction of the photosphere in the E-shell increases (panel c of Figure \ref{fig:SN2024ggiFigureSnapshots}), until the entire photosphere is at the E-shell boundary (panel d of Figure \ref{fig:SN2024ggiFigureSnapshots}).  
The line of sight in these figures is along $z$-axis, perpendicular to the long axis of the E-shell. 

For the evolution of the photospheric radius of the S-shell, $R_{\rm ph,S}(t)$, we use equations from \cite{Liu2018}, who derive the relation 
\begin{equation}
    R_{\rm ph,S} (t) = v_{\rm S} t - \frac{2}{3}\lambda(t) ,
    \label{eq:RphS}
\end{equation} 
where $v$ is the material velocity for homologous expansion, and $\lambda = \frac{3}{2}Bt^3$ is a recession term (and B is the recession coefficient) due to the thinning of the ejecta (becoming optically thin) in the case where the density is uniform in the sphere and decreases as $t^{-3}$.
We can find an exact analytical form for the photospheric radius, in the uniform profile case, in terms of the physical parameters: Ejecta Mass ($M_{ej}$), Kinetic Energy ($E_K$), and Opacity ($\kappa$):
\begin{equation}
\begin{split}
R_{\rm ph,S}(t) = & \left[  \left(\frac{10 E_K}{3 M_{\rm ej}}\right)^{1/2} \right] t  
 \\ & -
\left[ \frac{8 \pi}{9 \kappa M_{\rm ej}} 
  \left(\frac{10 E_K}{3 M_{\rm ej}}\right)^{3/2} \right] t^3 
    = v_{\rm S} t - Bt^3  .
\label{eq:HomRph}
\end{split}
\end{equation}
Equation (\ref{eq:HomRph}) reproduces the results of  \cite{Liu2018} for a spherical shell. In this study, we will not try to fit the explosion energy and ejecta mass, but will give only one possible set of values in Section \ref{sec:Photosphere}. 

For the E-shell, we take an ellipsoid with axes $a(t)=v_a t$  and $b(t)=c(t)=v_b t < a(t)$, for all $t$.  We define 
\begin{equation}
    R_{\rm ph,E} (t) = \sqrt{a(t)b(t)} = t \sqrt{v_av_b} ,
\label{eq:RphE}
\end{equation} 
so that the effective projected photospheric area for an observer perpendicular to the long axis when only the E-shell contributes is $\pi R^2_{\rm ph,E}$. This allows us to define an equivalent velocity of the outer boundary of the E-shell, $v_{\rm eq} \equiv \sqrt{v_a v_b}$, so that $R_{\rm ph,E} (t) =v_{\rm eq} t$.

We aim to reproduce the photospheric radius derived from observations that do not resolve the ejecta. The radius is derived from the luminosity and temperature, assuming a spherical photosphere. 
Before the E-shell starts to contribute, i.e., $R_{\rm ph,S}(t) \ge a(t)$, the photospheric radius is that given by equation (\ref{eq:RphS}) or (\ref{eq:HomRph}), $R_{\rm ph}= R_{\rm ph,S}$. 
 After the S-shell ceases to contribute, i.e. it is fully contained inside the E-shell, $R_{\rm ph,S}(t) \le b(t)$, the photospheric radius is given by equation (\ref{eq:RphE}): $R_{\rm ph}= R_{\rm ph,E}$. 

To calculate the observed photospheric radius when both contribute, we proceed as follows.  
We calculate the projected area that each shell contributes to the photosphere when $b(t) < R_{\rm ph,S}(t) < a(t)$, i.e., both contribute to the photosphere (see Figure \ref{fig:SN2024ggiFigureVisualization}). 
The projected area contribution of the S-shell to the photosphere is 
\begin{equation}
\begin{split}
    A_{\rm S}(t) = &2 x_{\rm int}(t) y_{\rm int}(t) \\&
    + 2 R^2_{\rm ph,S}(t) \arcsin\left[\frac{x_{\rm int}(t)}{R_{\rm ph,S}(t)}\right]  ,
\end{split}
\end{equation}
and that of the E-shell is 
\begin{equation}
\begin{split}
    A_{\rm E}(t) = &\pi a(t)b(t) - 2x_{\rm int} (t)y_{\rm int}(t) \\&
    - 2 a(t)b(t) \arcsin\left[\frac{x_{\rm int}(t)}{a(t)}\right]
\end{split}
\end{equation}
where 
\begin{equation}
    x_{\rm int} = a \sqrt{\frac{R_{\rm ph,S}^2 - b^2}{a^2 - b^2}}, \quad 
    y_{\rm int} = b \sqrt{\frac{a^2 - R_{\rm ph,S}^2}{a^2 - b^2}} ,
\end{equation} 
for $b(t) < R_{\rm ph,S}(t) < a(t)$.

 When the effective temperatures of the two shells are the same, $T_{\rm S} =T_{\rm E}$, the observationally derived photospheric radius is 
 \begin{equation}
    R_{\rm ph} (t) = \sqrt{\frac{A_{\rm S}(t) + A_{\rm E}(t)}{\pi} }, \quad {\rm for} \quad T_{\rm S} =T_{\rm E}  .
\label{eq:RphSE}
\end{equation} 
 
The two shells may have somewhat different temperatures. In the present toy model, we keep the temperature ratio constant in that case (although the temperatures vary substantially with time). The flux is proportional to the effective temperature of each shell to the fourth power, such that the luminosity that each shell contributes is $L_{\rm S}=A_{\rm S} \sigma T^4_{\rm S}$, and $L_{\rm E}=A_{\rm E} \sigma T^4_{\rm E}$, where $\sigma$ is the Stefan-Boltzmann constant. When the two temperatures are close, observations will deduce one effective temperature, which we take to be 
\begin{equation} \label{eq:teff}
    T_{\rm eff}^4 = \frac{L_{\rm E}T^4_{\rm E} + L_{\rm S} T^4_{\rm S}}{L_{\rm E} + L_{\rm S}} .  
\end{equation}
The derived photospheric radius is 
\begin{equation}
    R_{\rm ph} = \sqrt 
      {\frac{L_{\rm E} + L_{\rm S}}
      {\pi \sigma T^4_{\rm eff}} } 
      = \sqrt{  \frac{(A_{\rm E} T^4_{\rm E} + A_{\rm S} T^4_{\rm S})^2} 
                     {\pi (A_{\rm E} T_{\rm E}^8 + A_{\rm S} T_s^8)}} .
    \label{eq:RphSET}
\end{equation}

It can be shown that, for a given geometry of the two shells, the maximum radius  $R_{\rm ph}$ in the transition phase is achieved when the temperatures are equal, i.e., $T_e=T_s$. The corresponding photospheric radius in this case is given by equation \ref{eq:RphSE}.

\section{Identifying photospheric shells in SN 2024ggi}
\label{sec:Photosphere}

SN~2024ggi is a CCSN that attracted immediate attention because it is relatively close and has a circumstellar matter ( e.g., \citealt{ChenXetal2024, JacobsonGalanetal2024, Hongetal2024, Pessietal2024, Shresthaetal2024, Xiangetal2024, Aryanetal2025, Baronetal2025, Bostroemetal2025, Ertinietal2025, Ferrarietal2025, Huetal2025, Hueichapanetal2025, Meraetal2025}).
In this study, we examine only the evolution of the observationally determined photospheric radius of the ejecta (e.g., \citealt{ZhangJetal2024, ChenTWetal2025}). Observations do not resolve SN 2024ggi, and the radius of the photosphere is calculated under the assumption of a spherically symmetric ejecta. We will use the non-spherical model that we described in Section \ref{sec:Method} to fit the photospheric radius evolution of SN 2024ggi.  

\begin{figure}[th]
\begin{center}
\includegraphics[trim=0.0cm 0.00cm 0.0cm 0.0cm ,clip, scale=0.29]{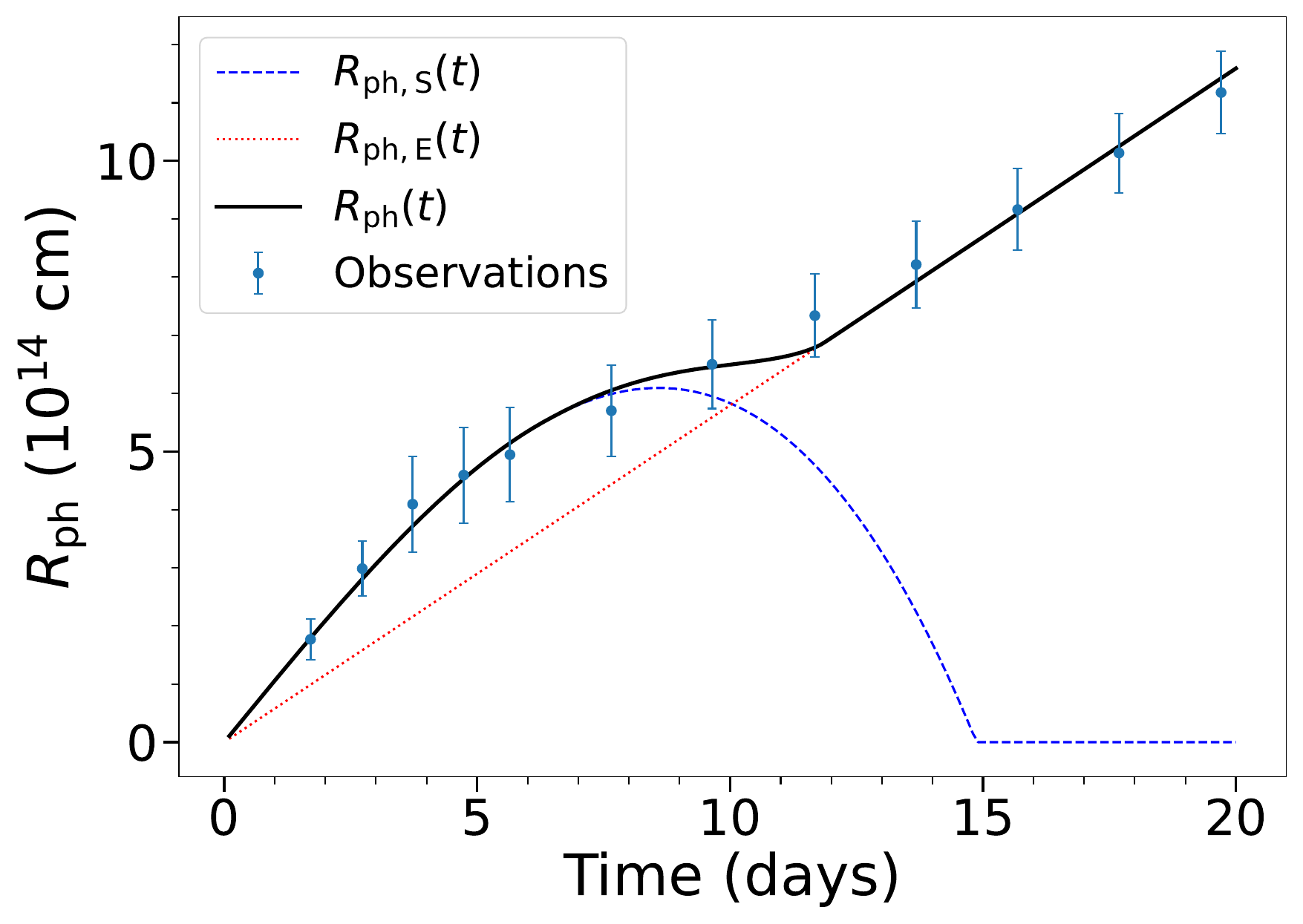}
\caption{The thick black line represents the photospheric radius in our fiducial toy model (Tables \ref{tab:FiducialModelS} and \ref{tab:FiducialModelE}), according to equation (\ref{eq:RphSE}). The observations are from \cite{ChenTWetal2025}. The dashed blue line is the photospheric radius of the S-shell according to equation (\ref{eq:RphS}), and the dotted red line is the photospheric radius of the E-shell according to equation (\ref{eq:RphE}), both in the fiducial model. }
\label{fig:SN2024ggiFigurePhotosphereFiducial}
\end{center}
\end{figure}

In Figure \ref{fig:SN2024ggiFigurePhotosphereFiducial}, we present the photospheric radius of the ejecta, based on the observational study by \cite{ChenTWetal2025}. We also present the evolution of the radius according to our fiducial toy model. When only the S-shell contributes to the photosphere,  at early times $R_{\rm ph,S}(t) > v_a t$, the theoretical radius in the fiducial model is given by equation (\ref{eq:RphS}) with $v_S=12,323 \km \s^{-1}$ and $B=7.46 \times 10^{-9} \km \s^{-1}$. When only the E-shell contributes to the photosphere, at late times $R_{\rm ph,S}(t) < v_b t$, it is given by equation (\ref{eq:RphE}) with $v_a=10,000 \km \s^{-1}$ and $v_b=4500 \km \s^{-1}$. In the fiducial model, the effective temperatures of the two shells are equal, $T_{\rm E} =T_{\rm S}$; thus, equation (\ref{eq:RphSE}) yields the radius during the transition phase when both shells contribute to the photosphere, i.e., when $v_b t < R_{\rm ph,S}(t) < v_a t$. The dashed blue line is equation (\ref{eq:RphS}) and the dotted red line is equation (\ref{eq:RphE}). 

We summarize the parameters of the fiducial model in Tables \ref{tab:FiducialModelS} and \ref{tab:FiducialModelE}. For the S-shell, we obtain a fitting with the parameters $v_{\rm S}$ and $B$. The values of the possible fitting set of values for the opacity, mass, and energy of the spherical shell that appear in equation (\ref{eq:HomRph}) are degenerate. Namely, other sets of values are possible. The key point is that the S-shell is of low mass. 
\begin{table}[h]
\begin{center}
\caption{Fiducial model parameters for the S-shell}
\begin{tabular}{|l|c|c|}
\hline
\textbf{Parameter} & \textbf{Value} \\
\hline
S-shell Mass           & $M_{\rm ej} = 0.016  \rmModot$  \\
S-shell Kinetic Energy & $E_k = 1.45 \times 10^{49} \erg$ \\
S-shell Opacity               & $\kappa = 0.22  \cm^{2} \g^{-1}$ \\

Material velocity (by eq. \ref{eq:HomRph}) & $v_{\rm S} = 12,323 \km \s^{-1}$ \\
Recession coefficient (by eq. \ref{eq:HomRph}) & $B = 7.46 \times 10^{-9} \km \s^{-3}$ \\
\hline
\end{tabular}
\label{tab:FiducialModelS}
\end{center}
\begin{flushleft}
\small 
Notes: Properties of the S-shell in the fiducial model. The fiducial model assumes a uniform density in the S-shell. Note that the set of values for the opacity, mass, and energy of the spherical shell that appear in equation (\ref{eq:HomRph}) is degenerate. Namely, other sets of values are possible. We actually determine the values of $v_S$ and $B$ by fitting to observations.   \end{flushleft}
\end{table}
\begin{table}[h]
\begin{center}
\caption{Fiducial model parameters for the E-shell}
\begin{tabular}{|l|c|}
\hline
\textbf{Parameter} & \textbf{Value} \\
\hline
Long-axis  velocity & $v_a = 10,000 \km \s^{-1}$  \\
Short-axis velocity & $v_b = 4500 \km \s^{-1}$  \\
Equivalent velocity & $v_{\rm eq} = \sqrt{v_a v_b} = 6708 \km \s^{-1}$ \\
\hline  
\end{tabular}
\label{tab:FiducialModelE}
\end{center}
\begin{flushleft}
\small 
Notes: Given are the velocities of the front of the E-shell along the long and short axes, and the equivalent velocity in the fiducial model. In the fiducial model, the two temperatures are equal, $T_{\rm E}=T_{\rm S}$. The E-shell is dense and massive, and it remains optically thick throughout the simulated period. Namely, its photosphere is at its front.
\end{flushleft}
\end{table}

\begin{figure}[th]
\begin{center}
\includegraphics[trim=0.0cm 0.00cm 0.0cm 0.0cm ,clip, scale=0.29]{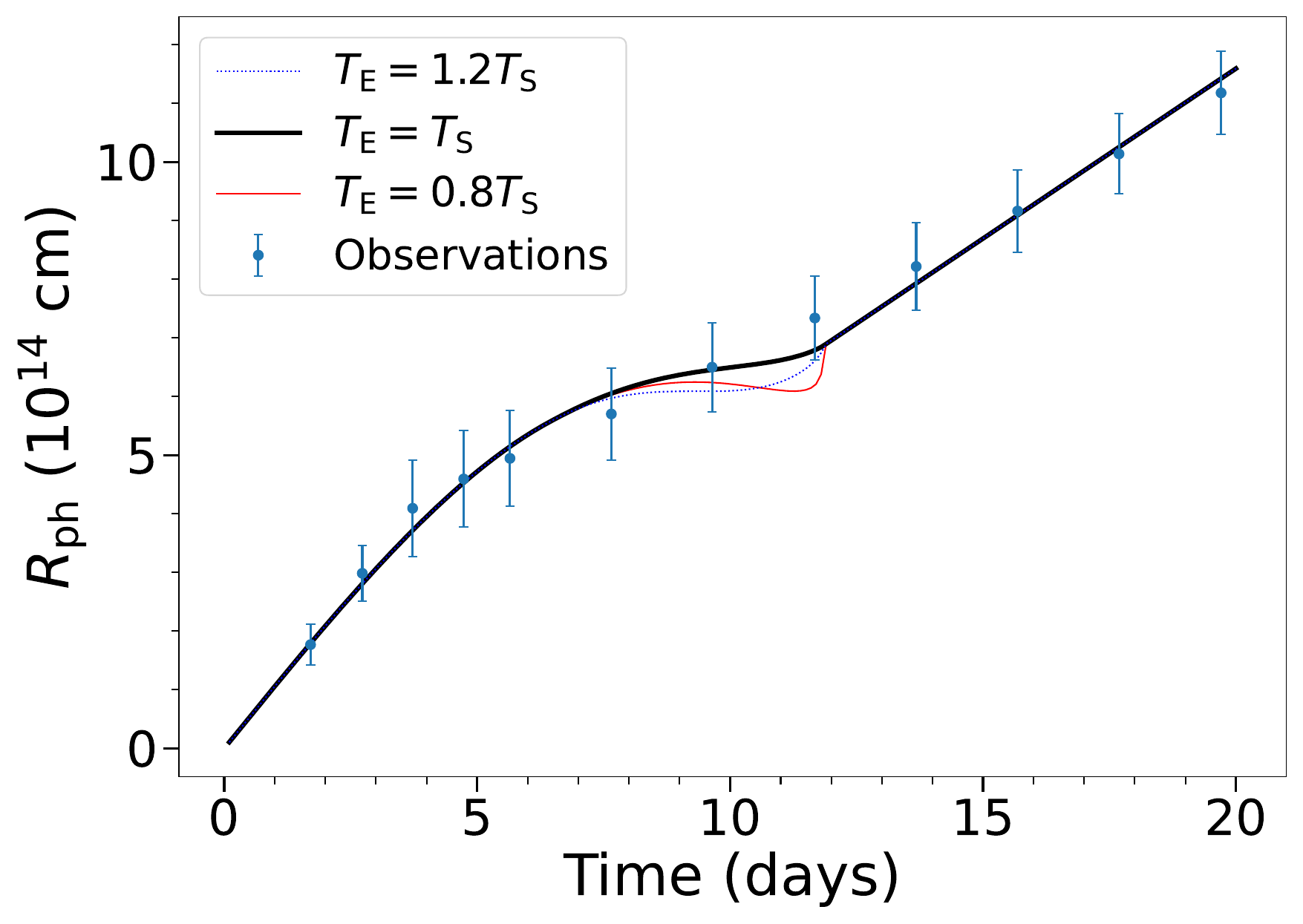}
\caption{The photospheric radius evolution in three models that differ in their effective temperature ratio, as indicated in the inset. When the two temperatures are unequal, the photospheric radius is according to equation (\ref{eq:RphSET}). All models have the same geometry as in the fiducial models (Figure \ref{fig:SN2024ggiFigureSnapshots}). The temperature ratio affects the photospheric radius only during the transition phase, i.e., when both shells contribute to the photosphere. The thick black line represents the fiducial model. For a given geometry, the case with equal temperatures yields the largest calculated radius in the transition phase.}
\label{fig:SN2024ggiFigurePhotosphereVarTemperatures}
\end{center}
\end{figure}

The key observation for our modeling is the transition from concave, $d^2 R_{\rm ph} /dt^2 < 0$,  to convex, $d^2 R_{\rm ph} /dt^2 > 0$, behavior. A single shell that becomes increasingly transparent forms a concave function, namely, one with a decreasing slope. The photospheric radius evolution of SN 2024ggi, as calculated by \cite{ChenTWetal2025}, has a concave behavior in the first week, more or less. However, over the following days, the slope no longer decreases, and after about two weeks, it increases: $d^2 R_{\rm ph}/dt^2 > 0$. A single-shell ejecta has difficulties explaining this behavior. This is our motivation to construct the two-shell toy model for SN 2024ggi. The specific model of a spherical shell and an elongated shell is based on observations of some supernova remnants.  However, as we discuss in Section \ref{sec:Summary}, other multi-shell structures are possible,  
    
When the effective temperatures of the two shells differ, the photospheric radius during the transition phase is according to equation (\ref{eq:RphSET}). For a given geometry of the two shells, equal temperatures give the largest radius. Figure \ref{fig:SN2024ggiFigurePhotosphereVarTemperatures} presents two cases with unequal temperatures, along with the fiducial toy model. In our toy model, the fit to the observations degrades with unequal temperatures. 

\begin{figure}[th]
\begin{center}
\includegraphics[trim=0.0cm 0.00cm 0.0cm 0.0cm ,clip, scale=0.29]{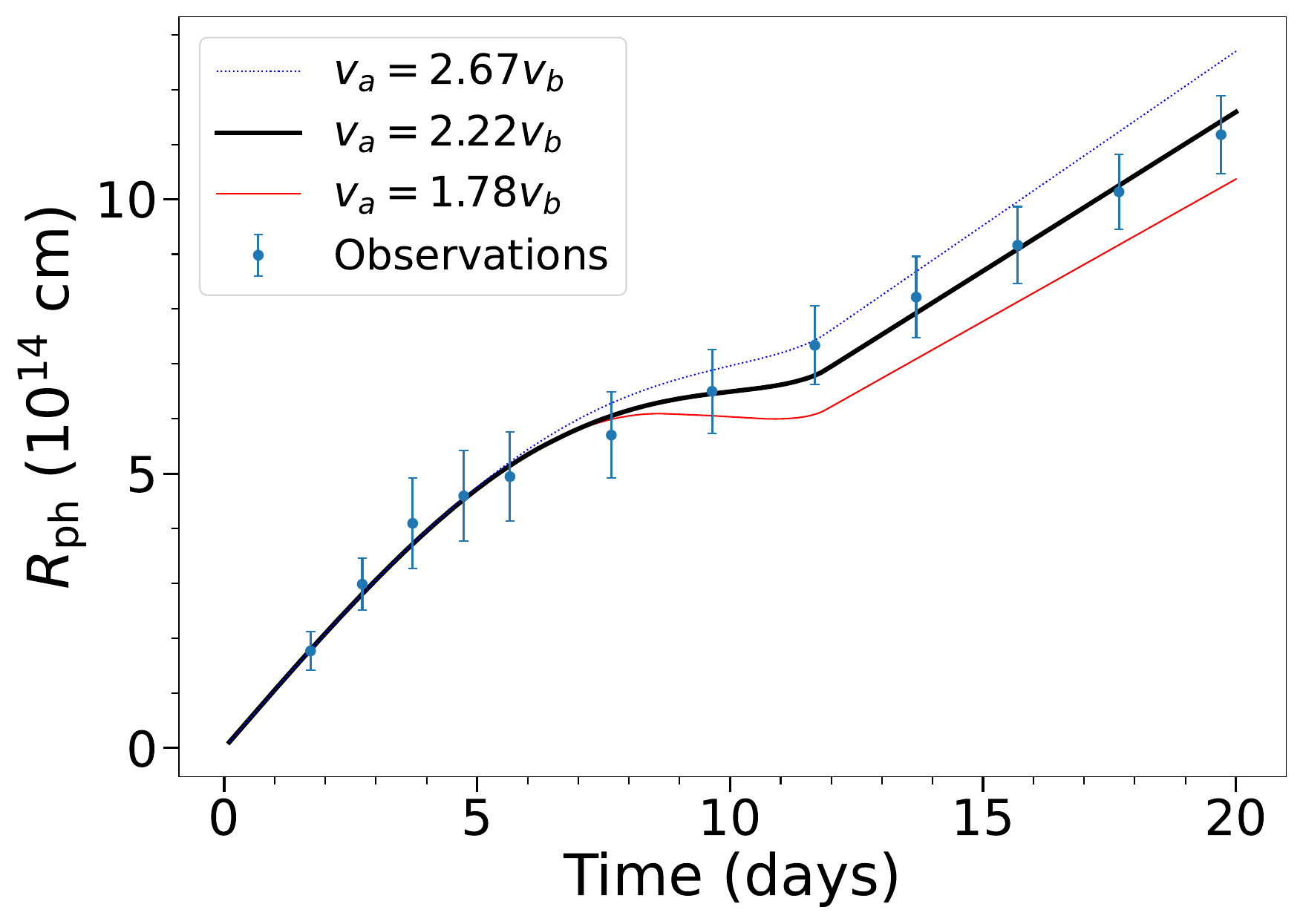}
\caption{The photospheric radius evolution in three models that differ in the aspect ratio of the E-shell, i.e., $a/b=v_a/v_b$. The three cases have $v_b=4500 \km \s^{-1}$ as in the
fiducial mode, and differ in the velocity of the long axis: $v_a=8000 \km \s^{-1}$ (lower line), $v_a=10,000 \km \s^{-1}$ (the fiducial model in thick black line), and $v_a=12,000  \km \s^{-1}$ (upper line). 
}
\label{fig:SN2024ggiFigurePhotosphereVarVelocities}
\end{center}
\end{figure}

In Figure \ref{fig:SN2024ggiFigurePhotosphereVarVelocities}, we present cases that differ in the long-axis velocity $v_a$. In addition to the fiducial toy model with $v_a=10,000 \km \s^{-1}$, we present cases with slower, $v_a=8000 \km \s^{-1}$ (lower line) and faster long-axis velocities, $v_a=12,000 \km \s^{-1}$ (upper line). As $v_a$ increases, the time at which the E-shell starts to contribute is earlier. However, at later times, these two other cases do not align with the observations. 

Several comments are in place here. 
(1) The velocity of the S-shell of $v_{\rm S} = 12,323  \km \s^{-1}$ in the fiducial model is compatible with the observed velocity determined by Doppler shift in the first several days (e.g., \citealt{JacobsonGalanetal2024}).  
(2) We consider two shells, each covering a solid angle of $4 \pi$. However, shells may cover a large fraction of the viewing angle, but not full $4 \pi$ around the center of the explosion. The low mass of the S-shell might suggest that it is a low-mass partial shell. Such a partial low-mass shell might appear as rims in the lobes or ears of CCSNRs (Section \ref{sec:Summary} and \citealt{SokerShiran2025}). Namely, it is possible that the S-shell (first one) is not a complete sphere. 
(3) The large polarization from early times \citep{YangYietal2025} in SN 2024ggi might support that the S-shell is also not spherical. We discuss this in Section \ref{sec:Summary}.  
(4) The line of sight we consider is perpendicular to the symmetry axis of the E-shell. Observers along and near the symmetry axis of the E-shell will not observe the behavior under study. Nonetheless, because of the axial symmetry around the long axis, half of the observers in a random distribution will be less than $30^\circ$ from the direction we study. Therefore, a single direction is sufficient for the present toy model.   
(5) Our toy model neglects the circumstellar matter. Future studies that develop more sophisticated models should incorporate circumstellar matter.  

Overall, our results indicate the need for multi-shell ejecta in SN 2024ggi. 

\section{Summary} 
\label{sec:Summary}

We built a simple toy model composed of two shells to explain the evolution of the photospheric radius of SN 2024ggi. Motivated by the morphology of some CCSNRs (Figure \ref{fig:SN2024ggiFigureCCSNRs}), we composed the ejecta from a spherical component, the S-shell, and a prolate ellipsoid, the E-shell, that has a slower expansion velocity (Figure \ref{fig:SN2024ggiFigureVisualization}). At early times, the faster S-shell forms the photosphere (see panel a in Figure \ref{fig:SN2024ggiFigureSnapshots}). As the S-shell expands, the photosphere moves inward in mass coordinate, although it continues to grow. Within days, the photosphere of the S-shell moves into the edges of the E-shell, and the E-shell starts to contribute to the photosphere (panels b and c of Figure \ref{fig:SN2024ggiFigureSnapshots}). Eventually, only the E-shell contributes to the photosphere (panel d). The E-shell is massive and dense, and the photosphere lies at its outer edge during the period covered in this study. From this behavior, we can calculate the photospheric radius, as observations would deduce under the assumption of spherical ejecta only: equation (\ref{eq:RphSE}) for equal temperatures and equation (\ref{eq:RphSET}) for different temperatures of the two shells. 

We find that this simple toy model reproduces the observed photospheric radius reported in \cite{ChenTWetal2025} quite well. Figure \ref{fig:SN2024ggiFigurePhotosphereFiducial} presents the calculated photospheric radius in our fiducial model (Tables \ref{tab:FiducialModelS} and \ref{tab:FiducialModelE}), along with the contributions from the S-shell and E-shell. In Figure \ref{fig:SN2024ggiFigurePhotosphereVarTemperatures}, we present cases of unequal temperatures, and in Figure \ref{fig:SN2024ggiFigurePhotosphereVarVelocities}, cases of different elongation of the E-shell; these figures demonstrate the role of the relative temperatures and the E-shell elongation in our toy model. We conclude that a two-shell model of the ejecta, with the slower shell more elongated and more massive than the faster shell, can reproduce the photospheric radius of SN 2024ggi.  
   
We note that \cite{YangYietal2025} measured polarization in SN 2024ggi, which varies with time. They conclude that the physical mechanism driving the explosion of massive stars exhibits well-defined axial symmetry and operates on large scales. This conclusion is compatible with the morphologies of many CCSNRs whose shaping is attributed to jets in the JJEM (see earlier papers in Section \ref{sec:intro}). From their spectropolarimetry, \cite{YangYietal2025} concluded that SN 2024ggi has a moderately aspherical explosion with a well-defined symmetry axis shared by the prompt shock-breakout emission and the SN ejecta. Our very simple toy model has an initial spherical shell. However, this toy model can be extended to include two elongated shells (ellipsoidal or other elongated shapes). In our study of SN 2023ixf \citep{SokerShiran2025}, we presented images of two CCSNRs with elongated shells oriented in the same direction. These are SNR W44, which has two elongated shells aligned along the same axis, and SNR G0.9+0.1, which has a large protrusion (an `ear') with two prominent rims, also aligned along the same axis. Both CCSNRs have jet-shaped morphologies that studies have attributed to the JJEM (\citealt{Soker2024W44, Soker2025G0901}).
The very low mass and energy that we estimate for the S-shell (Table \ref{tab:FiducialModelS}) might point to an external `rim' that covers a large fraction of the area towards the observer as a possibility for the S-shell structure. Building a more sophisticated model with multiple shells that accounts for polarization and the photospheric radius of SN 2024ggi and other CCSNe is a topic for future study. Such models should include multiple non-spherical shells, consider more complex structures, such as lobes and ears, and eventually incorporate radiation-hydrodynamics simulations. 
While a full radiation-hydrodynamics simulation is beyond the scope of this work, our simple toy model establishes the minimal structural complexity required to explain the data.
  
In \cite{SokerShiran2025}, we identified two or even three shells in the photospheric radius evolution of SN 2023ixf as calculated by \cite{Zimmermanetal2024}. In addition to the present study, multiple-shell ejecta appear to be common in CCSNe.  
The structure of multiple-shell ejecta is a prediction of the JJEM, in which up to several pairs of energetic jets participate in the explosion (e.g., \citealt{Braudoetal2025}; more low-energy pairs of jets might also exist). As discussed in this study and that on SN 2023ixf, the multiple-shell ejecta are consistent with several CCSNRs with jet-shaped morphologies that studies have attributed to the JJEM. Indeed, CCSNR morphologies strongly support the JJEM and severely challenge the neutrino-driven explosion mechanism.  Our studies of SN 2023ixf and SN 2024ggi show that the photospheric phase of CCSNe can also be used to support or challenge CCSN explosion mechanisms; our two studies of multiple-shell ejecta, and studies of polarization (e.g., \citealt{YangYietal2025}), are better compatible with the JJEM than with the neutrino-driven mechanism. Our study supports the growing evidence that the JJEM is the primary explosion mechanism of CCSNe.    

\section*{Acknowledgements}

NS thanks the Charles Wolfson Academic Chair at the Technion for the support.


%
\bibliography{reference}{}

@ARTICLE{Soker2024PNSN,
       author = {{Soker}, Noam},
        title = "{Planetary Nebula Morphologies Indicate a Jet-Driven Explosion of SN 1987A and Other Core-Collapse Supernovae}",
      journal = {Galaxies},
         year = 2024,
        month = jun,
       volume = {12},
       number = {3},
          eid = {29},
        pages = {29},
          doi = {10.3390/galaxies12030029},
archivePrefix = {arXiv},
       eprint = {2404.14843},
 primaryClass = {astro-ph.HE},
       adsurl = {https://ui.adsabs.harvard.edu/abs/2024Galax..12...29S}
}

@ARTICLE{Soker2022nu,
       author = {{Soker}, Noam},
        title = "{Boosting Jittering Jets by Neutrino Heating in Core Collapse Supernovae}",
      journal = {Research in Astronomy and Astrophysics},
         year = 2022,
        month = sep,
       volume = {22},
       number = {9},
          eid = {095007},
        pages = {095007},
          doi = {10.1088/1674-4527/ac7cbc},
archivePrefix = {arXiv},
       eprint = {2202.05556},
 primaryClass = {astro-ph.HE},
       adsurl = {https://ui.adsabs.harvard.edu/abs/2022RAA....22i5007S}
}

@ARTICLE{Nakamuraetal2025,
       author = {{Nakamura}, Ko and {Takiwaki}, Tomoya and {Matsumoto}, Jin and {Kotake}, Kei},
        title = "{Three-dimensional magnetohydrodynamic simulations of core-collapse supernovae - I. Hydrodynamic evolution and protoneutron star properties}",
      journal = {\mnras},
         year = 2025,
        month = jan,
       volume = {536},
       number = {1},
        pages = {280-294},
          doi = {10.1093/mnras/stae2611},
archivePrefix = {arXiv},
       eprint = {2405.08367},
 primaryClass = {astro-ph.HE},
       adsurl = {https://ui.adsabs.harvard.edu/abs/2025MNRAS.536..280N}
}

@ARTICLE{Gaensleretal1998,
       author = {{Gaensler}, B.~M. and {Green}, A.~J. and {Manchester}, R.~N.},
        title = "{G309.2-00.6 and jets in supernova remnants}",
      journal = {\mnras},
         year = 1998,
        month = sep,
       volume = {299},
       number = {3},
        pages = {812-824},
          doi = {10.1046/j.1365-8711.1998.01814.x},
archivePrefix = {arXiv},
       eprint = {astro-ph/9805163},
 primaryClass = {astro-ph},
       adsurl = {https://ui.adsabs.harvard.edu/abs/1998MNRAS.299..812G}
}

@ARTICLE{ShishkinKayeSoker2024,
       author = {{Shishkin}, Dmitry and {Kaye}, Roy and {Soker}, Noam},
        title = "{Identifying Jittering Jet-shaped Ejecta in the Cygnus Loop Supernova Remnant}",
      journal = {\apj},
         year = 2024,
        month = nov,
       volume = {975},
       number = {2},
          eid = {281},
        pages = {281},
          doi = {10.3847/1538-4357/ad8138},
archivePrefix = {arXiv},
       eprint = {2408.11014},
 primaryClass = {astro-ph.HE},
       adsurl = {https://ui.adsabs.harvard.edu/abs/2024ApJ...975..281S}
}

@ARTICLE{Kumar2025,
       author = {{Kumar}, Amit},
        title = "{Insights from modelling magnetar-driven light curves of stripped-envelope supernovae}",
      journal = {\na},
         year = 2025,
        month = may,
       volume = {116},
          eid = {102346},
        pages = {102346},
          doi = {10.1016/j.newast.2024.102346},
archivePrefix = {arXiv},
       eprint = {2412.09357},
 primaryClass = {astro-ph.HE},
       adsurl = {https://ui.adsabs.harvard.edu/abs/2025NewA..11602346K}
}

@ARTICLE{Soker2025Learning,
       author = {{Soker}, Noam},
        title = "{Learning from core-collapse supernova remnants on the explosion mechanism}",
      journal = {\na},
         year = 2025,
        month = dec,
       volume = {121},
          eid = {102453},
        pages = {102453},
          doi = {10.1016/j.newast.2025.102453},
archivePrefix = {arXiv},
       eprint = {2409.13657},
 primaryClass = {astro-ph.HE},
       adsurl = {https://ui.adsabs.harvard.edu/abs/2025NewA..12102453S}
}

@ARTICLE{Bocciolietal2025,
       author = {{Boccioli}, Luca and {Vartanyan}, David and {O'Connor}, Evan P. and {Kasen}, Daniel},
        title = "{Neutrino heating in 1D, 2D, and 3D core-collapse supernovae: characterizing the explosion of high-compactness stars}",
      journal = {\mnras},
         year = 2025,
        month = jul,
       volume = {540},
       number = {4},
        pages = {3885-3905},
          doi = {10.1093/mnras/staf963},
archivePrefix = {arXiv},
       eprint = {2501.06784},
 primaryClass = {astro-ph.HE},
       adsurl = {https://ui.adsabs.harvard.edu/abs/2025MNRAS.540.3885B}
}

@ARTICLE{Soker2024UnivReview,
       author = {{Soker}, Noam},
        title = "{The Two Alternative Explosion Mechanisms of Core-Collapse Supernovae: 2024 Status Report}",
      journal = {Universe},
         year = 2024,
        month = dec,
       volume = {10},
       number = {12},
          eid = {458},
        pages = {458},
          doi = {10.3390/universe10120458},
archivePrefix = {arXiv},
       eprint = {2411.08555},
 primaryClass = {astro-ph.HE},
       adsurl = {https://ui.adsabs.harvard.edu/abs/2024Univ...10..458S}
}

@ARTICLE{Bearetal2025Puppis,
       author = {{Bear}, Ealeal and {Shishkin}, Dmitry and {Soker}, Noam},
        title = "{The Puppis A Supernova Remnant: An Early Jet-driven Neutron Star Kick followed by Jittering Jets}",
      journal = {Research in Astronomy and Astrophysics},
         year = 2025,
        month = apr,
       volume = {25},
       number = {4},
          eid = {045008},
        pages = {045008},
          doi = {10.1088/1674-4527/adc24e},
archivePrefix = {arXiv},
       eprint = {2409.11453},
 primaryClass = {astro-ph.HE},
       adsurl = {https://ui.adsabs.harvard.edu/abs/2025RAA....25d5008B}
}

@ARTICLE{Shibataetal2025,
       author = {{Shibata}, Masaru and {Fujibayashi}, Sho and {Wanajo}, Shinya and {Ioka}, Kunihito and {Lam}, Alan Tsz-Lok and {Sekiguchi}, Yuichiro},
        title = "{Self-consistent scenario for jet and stellar explosions in collapsar: General relativistic magnetohydrodynamics simulation with a dynamo}",
      journal = {\prd},
         year = 2025,
        month = jun,
       volume = {111},
       number = {12},
          eid = {123017},
        pages = {123017},
          doi = {10.1103/msy2-fwhx},
archivePrefix = {arXiv},
       eprint = {2502.02077},
 primaryClass = {astro-ph.HE},
       adsurl = {https://ui.adsabs.harvard.edu/abs/2025PhRvD.111l3017S}
}

@ARTICLE{Imashevaetal2025,
       author = {{Imasheva}, Liliya and {Janka}, Hans-Thomas and {Weiss}, Achim},
        title = "{Comparison of three methods for triggering core-collapse supernova explosions in spherical symmetry}",
      journal = {\mnras},
         year = 2025,
        month = jul,
       volume = {541},
       number = {1},
        pages = {116-134},
          doi = {10.1093/mnras/staf865},
archivePrefix = {arXiv},
       eprint = {2501.13172},
 primaryClass = {astro-ph.HE},
       adsurl = {https://ui.adsabs.harvard.edu/abs/2025MNRAS.541..116I}
}

@ARTICLE{SokerShishkin2025Vela,
       author = {{Soker}, Noam and {Shishkin}, Dmitry},
        title = "{The Vela Supernova Remnant: The Unique Morphological Features of Jittering Jets}",
      journal = {Research in Astronomy and Astrophysics},
         year = 2025,
        month = mar,
       volume = {25},
       number = {3},
          eid = {035008},
        pages = {035008},
          doi = {10.1088/1674-4527/adb4cc},
archivePrefix = {arXiv},
       eprint = {2409.02626},
 primaryClass = {astro-ph.HE},
       adsurl = {https://ui.adsabs.harvard.edu/abs/2025RAA....25c5008S}
}

@ARTICLE{Soker2024W44,
       author = {{Soker}, Noam},
        title = "{Identifying a Point-Symmetrical Morphology in the Core-Collapse Supernova Remnant W44}",
      journal = {Universe},
         year = 2024,
        month = dec,
       volume = {11},
       number = {1},
          eid = {4},
        pages = {4},
          doi = {10.3390/universe11010004},
archivePrefix = {arXiv},
       eprint = {2411.04654},
 primaryClass = {astro-ph.HE},
       adsurl = {https://ui.adsabs.harvard.edu/abs/2024Univ...11....4S}
}

@ARTICLE{BearSoker2025,
       author = {{Bear}, Ealeal and {Soker}, Noam},
        title = "{Identifying a point-symmetric morphology in supernova remnant Cassiopeia A: Explosion by jittering jets}",
      journal = {\na},
         year = 2025,
        month = jan,
       volume = {114},
          eid = {102307},
        pages = {102307},
          doi = {10.1016/j.newast.2024.102307},
archivePrefix = {arXiv},
       eprint = {2403.07625},
 primaryClass = {astro-ph.HE},
       adsurl = {https://ui.adsabs.harvard.edu/abs/2025NewA..11402307B}
}

@ARTICLE{Soker2025G0901,
       author = {{Soker}, Noam},
        title = "{The Morphology of Supernova Remnant G0.9+0.1 Implies Explosion by Jittering-jets}",
      journal = {Research in Astronomy and Astrophysics},
         year = 2025,
        month = nov,
       volume = {25},
       number = {11},
          eid = {115005},
        pages = {115005},
          doi = {10.1088/1674-4527/adfd23},
archivePrefix = {arXiv},
       eprint = {2504.11384},
 primaryClass = {astro-ph.HE},
       adsurl = {https://ui.adsabs.harvard.edu/abs/2025RAA....25k5005S}
}

@ARTICLE{Laplaceetal2025,
       author = {{Laplace}, E. and {Schneider}, F.~R.~N. and {Podsiadlowski}, Ph.},
        title = "{It's written in the massive stars: The role of stellar physics in the formation of black holes}",
      journal = {\aap},
         year = 2025,
        month = mar,
       volume = {695},
          eid = {A71},
        pages = {A71},
          doi = {10.1051/0004-6361/202451077},
archivePrefix = {arXiv},
       eprint = {2409.02058},
 primaryClass = {astro-ph.SR},
       adsurl = {https://ui.adsabs.harvard.edu/abs/2025A&A...695A..71L}
}

@ARTICLE{Huangetal2025,
       author = {{Huang}, Xu-Run and {Zha}, Shuai and {Chu}, Ming-chung and {O'Connor}, Evan P. and {Chen}, Lie-Wen},
        title = "{Phase-transition-induced Collapse of Proto-compact Stars and Its Implication for Supernova Explosions}",
      journal = {\apj},
         year = 2025,
        month = feb,
       volume = {979},
       number = {2},
          eid = {151},
        pages = {151},
          doi = {10.3847/1538-4357/ada146},
archivePrefix = {arXiv},
       eprint = {2409.16189},
 primaryClass = {astro-ph.HE},
       adsurl = {https://ui.adsabs.harvard.edu/abs/2025ApJ...979..151H}
}

@ARTICLE{EggenbergerAndersenetal2025,
       author = {{Eggenberger Andersen}, Oliver and {O'Connor}, Evan and {Andresen}, Haakon and {da Silva Schneider}, Andr{\'e} and {Couch}, Sean M.},
        title = "{Black Hole Supernovae, Their Equation of State Dependence, and Ejecta Composition}",
      journal = {\apj},
         year = 2025,
        month = feb,
       volume = {980},
       number = {1},
          eid = {53},
        pages = {53},
          doi = {10.3847/1538-4357/ada899},
archivePrefix = {arXiv},
       eprint = {2411.11969},
 primaryClass = {astro-ph.HE},
       adsurl = {https://ui.adsabs.harvard.edu/abs/2025ApJ...980...53E}
}

@ARTICLE{Maunderetal2025,
       author = {{Maunder}, Thomas and {Callan}, Fionntan P. and {Sim}, Stuart A. and {Heger}, Alexander and {M{\"u}ller}, Bernhard},
        title = "{Synthetic Light Curves and Spectra for the Photospheric Phase of a 3D Stripped-Envelope Supernova Explosion Model}",
      journal = {arXiv e-prints},
         year = 2024,
        month = oct,
          eid = {arXiv:2410.20829},
        pages = {arXiv:2410.20829},
          doi = {10.48550/arXiv.2410.20829},
archivePrefix = {arXiv},
       eprint = {2410.20829},
 primaryClass = {astro-ph.HE},
       adsurl = {https://ui.adsabs.harvard.edu/abs/2024arXiv241020829M}
}

@ARTICLE{Mulleretal2025,
       author = {{M{\"u}ller}, Bernhard and {Heger}, Alexander and {Powell}, Jade},
        title = "{Minimum Neutron Star Mass in Neutrino-Driven Supernova Explosions}",
      journal = {\prl},
         year = 2025,
        month = feb,
       volume = {134},
       number = {7},
          eid = {071403},
        pages = {071403},
          doi = {10.1103/PhysRevLett.134.071403},
archivePrefix = {arXiv},
       eprint = {2407.08407},
 primaryClass = {astro-ph.HE},
       adsurl = {https://ui.adsabs.harvard.edu/abs/2025PhRvL.134g1403M}
}

@ARTICLE{WangBurrows2025,
       author = {{Wang}, Tianshu and {Burrows}, Adam},
        title = "{The Effect of the Fast-flavor Instability on Core-collapse Supernova Models}",
      journal = {\apj},
         year = 2025,
        month = jun,
       volume = {986},
       number = {2},
          eid = {153},
        pages = {153},
          doi = {10.3847/1538-4357/add889},
archivePrefix = {arXiv},
       eprint = {2503.04896},
 primaryClass = {astro-ph.HE},
       adsurl = {https://ui.adsabs.harvard.edu/abs/2025ApJ...986..153W}
}

@ARTICLE{SykesMuller2025,
       author = {{Sykes}, Bailey and {M{\"u}ller}, Bernhard},
        title = "{Long-time 3D supernova simulations of nonrotating progenitors with magnetic fields}",
      journal = {\prd},
         year = 2025,
        month = mar,
       volume = {111},
       number = {6},
          eid = {063042},
        pages = {063042},
          doi = {10.1103/PhysRevD.111.063042},
archivePrefix = {arXiv},
       eprint = {2412.01155},
 primaryClass = {astro-ph.HE},
       adsurl = {https://ui.adsabs.harvard.edu/abs/2025PhRvD.111f3042S}
}

@ARTICLE{Janka2025,
       author = {{Janka}, H. -Thomas},
        title = "{Long-Term Multidimensional Models of Core-Collapse Supernovae: Progress and Challenges}",
      journal = {arXiv e-prints},
         year = 2025,
        month = feb,
          eid = {arXiv:2502.14836},
        pages = {arXiv:2502.14836},
          doi = {10.48550/arXiv.2502.14836},
archivePrefix = {arXiv},
       eprint = {2502.14836},
 primaryClass = {astro-ph.HE},
       adsurl = {https://ui.adsabs.harvard.edu/abs/2025arXiv250214836J}
}

@ARTICLE{ParadisoCoughlin2025,
       author = {{Paradiso}, Daniel A. and {Coughlin}, Eric R.},
        title = "{Gotta Go Fast: A Generalization of the Escape Speed to Fluid-dynamical Explosions and Implications for Astrophysical Transients}",
      journal = {\apj},
         year = 2025,
        month = jun,
       volume = {985},
       number = {2},
          eid = {173},
        pages = {173},
          doi = {10.3847/1538-4357/adce6f},
archivePrefix = {arXiv},
       eprint = {2504.11527},
 primaryClass = {astro-ph.HE},
       adsurl = {https://ui.adsabs.harvard.edu/abs/2025ApJ...985..173P}
}

@ARTICLE{Braudoetal2025,
       author = {{Braudo}, Jessica and {Michaelis}, Amir and {Akashi}, Muhammad and {Soker}, Noam},
        title = "{Simulating the Shaping of Point-symmetric Structures in the Jittering Jets Explosion Mechanism}",
      journal = {\pasp},
         year = 2025,
        month = may,
       volume = {137},
       number = {5},
          eid = {054201},
        pages = {054201},
          doi = {10.1088/1538-3873/add08e},
archivePrefix = {arXiv},
       eprint = {2503.10326},
 primaryClass = {astro-ph.HE},
       adsurl = {https://ui.adsabs.harvard.edu/abs/2025PASP..137e4201B}
}

@ARTICLE{Bambaetal2025CasA,
       author = {{Bamba}, Aya and {Agarwal}, Manan and {Vink}, Jacco and {Plucinsky}, Paul and {Terada}, Yukikatsu and {Behar}, Ehud and {Katsuda}, Satoru and {Mori}, Koji and {Sawada}, Makoto and {Matsumoto}, Hironori and {Corrales}, Lia and {Foster}, Adam and {Fujimoto}, Shin-ichiro and {Gu}, Liyi and {Ichikawa}, Kazuhiro and {Matsunaga}, Kai and {Mizuno}, Tsunefumi and {Murakami}, Hiroshi and {Nakajima}, Hiroshi and {Sato}, Toshiki and {Sonoda}, Haruto and {Suzuki}, Shunsuke and {Tateishi}, Dai and {Uchida}, Hiroyuki and {Ichihashi}, Masahiro and {Nobukawa}, Kumiko and {Orlando}, Salvatore},
        title = "{Measuring the asymmetric expansion of the Fe ejecta of Cassiopeia A with XRISM/Resolve}",
      journal = {\pasj},
         year = 2025,
        month = may,
          doi = {10.1093/pasj/psaf041},
archivePrefix = {arXiv},
       eprint = {2504.03268},
 primaryClass = {astro-ph.HE},
       adsurl = {https://ui.adsabs.harvard.edu/abs/2025PASJ..tmp...58B}
}

@ARTICLE{Shishkinetal2025S147,
       author = {{Shishkin}, Dmitry and {Bear}, Ealeal and {Soker}, Noam},
        title = "{Natal kick by early-asymmetrical pairs of jets to the neutron star of supernova remnant S147}",
      journal = {arXiv e-prints},
         year = 2025,
        month = jun,
          eid = {arXiv:2506.21548},
        pages = {arXiv:2506.21548},
archivePrefix = {arXiv},
       eprint = {2506.21548},
 primaryClass = {astro-ph.HE},
       adsurl = {https://ui.adsabs.harvard.edu/abs/2025arXiv250621548S}
}

@ARTICLE{Vinketal2025,
       author = {{Vink}, Jacco and {Agarwal}, Manan and {Bamba}, Aya and {Gu}, Liyi and {Plucinsky}, Paul and {Behar}, Ehud and {Corrales}, Lia and {Foster}, Adam and {Fujimoto}, Shin-ichiro and {Ichihashi}, Masahiro and {Ichikawa}, Kazuhiro and {Katsuda}, Satoru and {Matsumoto}, Hironori and {Matsunaga}, Kai and {Mizuno}, Tsunefumi and {Mori}, Koji and {Murakami}, Hiroshi and {Nakajima}, Hiroshi and {Sato}, Toshiki and {Sawada}, Makoto and {Sonoda}, Haruto and {Suzuki}, Shunsuke and {Tateishi}, Dai and {Terada}, Yukikatsu and {Uchida}, Hiroyuki},
        title = "{Mapping Cassiopeia A's silicon/sulfur Doppler velocities with XRISM-Resolve}",
      journal = {arXiv e-prints},
         year = 2025,
        month = may,
          eid = {arXiv:2505.04691},
        pages = {arXiv:2505.04691},
          doi = {10.48550/arXiv.2505.04691},
archivePrefix = {arXiv},
       eprint = {2505.04691},
 primaryClass = {astro-ph.HE},
       adsurl = {https://ui.adsabs.harvard.edu/abs/2025arXiv250504691V}
}

@ARTICLE{Morietal2025,
       author = {{Mori}, Kanji and {Takiwaki}, Tomoya and {Kotake}, Kei and {Horiuchi}, Shunsaku},
        title = "{Three-dimensional core-collapse supernova models with phenomenological treatment of neutrino flavor conversions}",
      journal = {\pasj},
         year = 2025,
        month = apr,
       volume = {77},
       number = {2},
        pages = {L9-L15},
          doi = {10.1093/pasj/psaf007},
archivePrefix = {arXiv},
       eprint = {2501.15256},
 primaryClass = {astro-ph.HE},
       adsurl = {https://ui.adsabs.harvard.edu/abs/2025PASJ...77L...9M}
}

@ARTICLE{Soker2025Padova,
       author = {{Soker}, Noam},
        title = "{The primary role of jets in powering core-collapse supernovae}",
      journal = {Video Memorie della Societa Astronomica Italiana},
         year = 2025,
        month = apr,
       volume = {2},
          eid = {47},
        pages = {47},
          doi = {10.36116/VIDEOMEM_2.2025.47},
       adsurl = {https://ui.adsabs.harvard.edu/abs/2025VMSAI...2...47S}
}

@ARTICLE{Maltsevetal2025,
       author = {{Maltsev}, K. and {Schneider}, F.~R.~N. and {Mandel}, I. and {M{\"u}ller}, B. and {Heger}, A. and {R{\"o}pke}, F.~K. and {Laplace}, E.},
        title = "{Explodability criteria for the neutrino-driven supernova mechanism}",
      journal = {\aap},
         year = 2025,
        month = aug,
       volume = {700},
          eid = {A20},
        pages = {A20},
          doi = {10.1051/0004-6361/202554931},
archivePrefix = {arXiv},
       eprint = {2503.23856},
 primaryClass = {astro-ph.SR},
       adsurl = {https://ui.adsabs.harvard.edu/abs/2025A&A...700A..20M}
}

@ARTICLE{Orlandoetal20251987A,
       author = {{Orlando}, S. and {Miceli}, M. and {Ono}, M. and {Nagataki}, S. and {Aloy}, M. -A. and {Bocchino}, F. and {Gabler}, M. and {Giudici}, B. and {Giuffrida}, R. and {Greco}, E. and {La Malfa}, G. and {Lee}, S. -H. and {Obergaulinger}, M. and {Petruk}, O. and {Sapienza}, V. and {Ustamujic}, S. and {Weng}, J.},
        title = "{Tracing the ejecta structure of supernova 1987A: Insights and diagnostics from 3D magnetohydrodynamic simulations}",
      journal = {\aap},
         year = 2025,
        month = jul,
       volume = {699},
          eid = {A305},
        pages = {A305},
          doi = {10.1051/0004-6361/202554862},
archivePrefix = {arXiv},
       eprint = {2504.19896},
 primaryClass = {astro-ph.HE},
       adsurl = {https://ui.adsabs.harvard.edu/abs/2025A&A...699A.305O}
}

@ARTICLE{Janka2025Padova,
       author = {{Janka}, Thomas},
        title = "{Core-collapse Supernova Theory in 2025: Progress and Puzzles}",
      journal = {Video Memorie della Societa Astronomica Italiana},
         year = 2025,
        month = apr,
       volume = {2},
          eid = {46},
        pages = {46},
          doi = {10.36116/VIDEOMEM_2.2025.46},
       adsurl = {https://ui.adsabs.harvard.edu/abs/2025VMSAI...2...46J}
}

@ARTICLE{SokerAkashi2025,
       author = {{Soker}, Noam and {Akashi}, Muhammad},
        title = "{The explosion jets of the core-collapse supernova remnant Circinus X-1}",
      journal = {arXiv e-prints},
         year = 2025,
        month = aug,
          eid = {arXiv:2508.10843},
        pages = {arXiv:2508.10843},
          doi = {10.48550/arXiv.2508.10843},
archivePrefix = {arXiv},
       eprint = {2508.10843},
 primaryClass = {astro-ph.HE},
       adsurl = {https://ui.adsabs.harvard.edu/abs/2025arXiv250810843S}
}

@ARTICLE{WangShishkinSoker2025,
       author = {{Wang}, Nikki Yat Ning and {Shishkin}, Dmitry and {Soker}, Noam},
        title = "{Jittering jets in stripped-envelope core-collapse supernovae}",
      journal = {arXiv e-prints},
         year = 2025,
        month = oct,
          eid = {arXiv:2510.02203},
        pages = {arXiv:2510.02203},
          doi = {10.48550/arXiv.2510.02203},
archivePrefix = {arXiv},
       eprint = {2510.02203},
 primaryClass = {astro-ph.HE},
       adsurl = {https://ui.adsabs.harvard.edu/abs/2025arXiv251002203W}
}

@ARTICLE{Soker2025N132D,
       author = {{Soker}, Noam},
        title = "{Attributing the point symmetric structure of core-collapse supernova remnant N132D to the jittering jets explosion mechanism}",
      journal = {arXiv e-prints},
         year = 2025,
        month = jul,
          eid = {arXiv:2507.00757},
        pages = {arXiv:2507.00757},
          doi = {10.48550/arXiv.2507.00757},
archivePrefix = {arXiv},
       eprint = {2507.00757},
 primaryClass = {astro-ph.HE},
       adsurl = {https://ui.adsabs.harvard.edu/abs/2025arXiv250700757S}
}

@ARTICLE{SokerShishkin2025W49B,
       author = {{Soker}, Noam and {Shishkin}, Dmitry},
        title = "{The main jet axis of the W49B supernova remnant}",
      journal = {\pasa},
         year = 2025,
        month = apr,
       volume = {42},
          eid = {e048},
        pages = {e048},
          doi = {10.1017/pasa.2025.39},
archivePrefix = {arXiv},
       eprint = {2502.09543},
 primaryClass = {astro-ph.HE},
       adsurl = {https://ui.adsabs.harvard.edu/abs/2025PASA...42...48S}
}

@ARTICLE{Willcoxetal2025,
       author = {{Willcox}, R. and {Schneider}, F.~R.~N. and {Laplace}, E. and {Podsiadlowski}, Ph. and {Maltsev}, K. and {Mandel}, I. and {Marchant}, P. and {Sana}, H. and {Li}, T. and {Hertog}, T.},
        title = "{New gravitational-wave data support a bimodal black-hole mass distribution}",
      journal = {arXiv e-prints},
         year = 2025,
        month = aug,
          eid = {arXiv:2508.20787},
        pages = {arXiv:2508.20787},
          doi = {10.48550/arXiv.2508.20787},
archivePrefix = {arXiv},
       eprint = {2508.20787},
 primaryClass = {astro-ph.HE},
       adsurl = {https://ui.adsabs.harvard.edu/abs/2025arXiv250820787W}
}

@ARTICLE{Tsunaetal2025,
       author = {{Tsuna}, Daichi and {Fuller}, Jim and {Lu}, Wenbin},
        title = "{Fates of Rotating Supergiants from Stellar Mergers and the Landscape of Transients upon Core-collapse}",
      journal = {arXiv e-prints},
         year = 2025,
        month = aug,
          eid = {arXiv:2508.21116},
        pages = {arXiv:2508.21116},
archivePrefix = {arXiv},
       eprint = {2508.21116},
 primaryClass = {astro-ph.HE},
       adsurl = {https://ui.adsabs.harvard.edu/abs/2025arXiv250821116T}
}

@ARTICLE{Soker2025RCW89,
       author = {{Soker}, Noam},
        title = "{Attributing the supernova remnant RCW 89 to the jittering jets explosion mechanism}",
      journal = {arXiv e-prints},
         year = 2025,
        month = sep,
          eid = {arXiv:2509.04723},
        pages = {arXiv:2509.04723},
archivePrefix = {arXiv},
       eprint = {2509.04723},
 primaryClass = {astro-ph.HE},
       adsurl = {https://ui.adsabs.harvard.edu/abs/2025arXiv250904723S}
}

@ARTICLE{Mukazhanov2025,
       author = {{Mukazhanov}, Olzhas},
        title = "{Impact of rotation on the accretion of entropy perturbations in collapsing massive stars}",
      journal = {arXiv e-prints},
         year = 2025,
        month = sep,
          eid = {arXiv:2509.09419},
        pages = {arXiv:2509.09419},
archivePrefix = {arXiv},
       eprint = {2509.09419},
 primaryClass = {astro-ph.SR},
       adsurl = {https://ui.adsabs.harvard.edu/abs/2025arXiv250909419M}
}

@ARTICLE{Raffeltetal2025,
       author = {{Raffelt}, Georg G. and {Janka}, Hans-Thomas and {Fiorillo}, Damiano F.~G.},
        title = "{Neutrinos from core-collapse supernovae}",
      journal = {arXiv e-prints},
         year = 2025,
        month = sep,
          eid = {arXiv:2509.16306},
        pages = {arXiv:2509.16306},
archivePrefix = {arXiv},
       eprint = {2509.16306},
 primaryClass = {astro-ph.HE},
       adsurl = {https://ui.adsabs.harvard.edu/abs/2025arXiv250916306R}
}

@ARTICLE{Vartanyanetal2025,
       author = {{Vartanyan}, David and {Burrows}, Adam and {Teryoshin}, Lizzy and {Wang}, Tianshu and {Kasen}, Daniel and {Tsang}, Benny and {Coleman}, Matthew S.~B.},
        title = "{Simulated 3D $^{56}$Ni Distributions of Type IIp Supernovae}",
      journal = {arXiv e-prints},
         year = 2025,
        month = sep,
          eid = {arXiv:2509.16314},
        pages = {arXiv:2509.16314},
archivePrefix = {arXiv},
       eprint = {2509.16314},
 primaryClass = {astro-ph.HE},
       adsurl = {https://ui.adsabs.harvard.edu/abs/2025arXiv250916314V}
}

@ARTICLE{FangQetal2025,
       author = {{Fang}, Qiliang and {Nagakura}, Hiroki and {Moriya}, Takashi J.},
        title = "{Reconciling the Tension Between Light Curve Modeling of Type II Supernovae and Neutrino-Driven Core-Collapse Supernovae Models with Late-Phase Spectroscopy}",
      journal = {arXiv e-prints},
         year = 2025,
        month = sep,
          eid = {arXiv:2509.20675},
        pages = {arXiv:2509.20675},
archivePrefix = {arXiv},
       eprint = {2509.20675},
 primaryClass = {astro-ph.HE},
       adsurl = {https://ui.adsabs.harvard.edu/abs/2025arXiv250920675F}
}

@ARTICLE{Zimmermanetal2024,
       author = {{Zimmerman}, E.~A. and {Irani}, I. and {Chen}, P. and {Gal-Yam}, A. and {Schulze}, S. and {Perley}, D.~A. and {Sollerman}, J. and {Filippenko}, A.~V. and {Shenar}, T. and {Yaron}, O. and {Shahaf}, S. and {Bruch}, R.~J. and {Ofek}, E.~O. and {De Cia}, A. and {Brink}, T.~G. and {Yang}, Y. and {Vasylyev}, S.~S. and {Ben Ami}, S. and {Aubert}, M. and {Badash}, A. and {Bloom}, J.~S. and {Brown}, P.~J. and {De}, K. and {Dimitriadis}, G. and {Fransson}, C. and {Fremling}, C. and {Hinds}, K. and {Horesh}, A. and {Johansson}, J.~P. and {Kasliwal}, M.~M. and {Kulkarni}, S.~R. and {Kushnir}, D. and {Martin}, C. and {Matuzewski}, M. and {McGurk}, R.~C. and {Miller}, A.~A. and {Morag}, J. and {Neil}, J.~D. and {Nugent}, P.~E. and {Post}, R.~S. and {Prusinski}, N.~Z. and {Qin}, Y. and {Raichoor}, A. and {Riddle}, R. and {Rowe}, M. and {Rusholme}, B. and {Sfaradi}, I. and {Sjoberg}, K.~M. and {Soumagnac}, M. and {Stein}, R.~D. and {Strotjohann}, N.~L. and {Terwel}, J.~H. and {Wasserman}, T. and {Wise}, J. and {Wold}, A. and {Yan}, L. and {Zhang}, K.},
        title = "{The complex circumstellar environment of supernova 2023ixf}",
      journal = {\nat},
         year = 2024,
        month = mar,
       volume = {627},
       number = {8005},
        pages = {759-762},
          doi = {10.1038/s41586-024-07116-6},
archivePrefix = {arXiv},
       eprint = {2310.10727},
 primaryClass = {astro-ph.HE},
       adsurl = {https://ui.adsabs.harvard.edu/abs/2024Natur.627..759Z}
}

@ARTICLE{Soker2025Dust,
       author = {{Soker}, Noam},
        title = "{Jittering jets promote dust formation in core-collapse supernovae}",
      journal = {arXiv e-prints},
         year = 2025,
        month = sep,
          eid = {arXiv:2509.19264},
        pages = {arXiv:2509.19264},
          doi = {10.48550/arXiv.2509.19264},
archivePrefix = {arXiv},
       eprint = {2509.19264},
 primaryClass = {astro-ph.HE},
       adsurl = {https://ui.adsabs.harvard.edu/abs/2025arXiv250919264S}
}

@ARTICLE{BoccioliRoberti2025,
       author = {{Boccioli}, Luca and {Roberti}, Lorenzo},
        title = "{Explodability matters: how realistic neutrino-driven explosions change explosive nucleosynthesis yields}",
      journal = {arXiv e-prints},
         year = 2025,
        month = oct,
          eid = {arXiv:2510.16365},
        pages = {arXiv:2510.16365},
archivePrefix = {arXiv},
       eprint = {2510.16365},
 primaryClass = {astro-ph.HE},
       adsurl = {https://ui.adsabs.harvard.edu/abs/2025arXiv251016365B}
}

@ARTICLE{PowellMuller2025,
       author = {{Powell}, Jade and   {M{\"u}ller}, Bernhard},
        title = "{Impact of the nuclear equation of state on the explodability of massive stars}",
      journal = {arXiv e-prints},
         year = 2025,
        month = Oct,
      eid = {arXiv:2510.20076},
        pages = {arXiv:2510.20076},
archivePrefix = {arXiv},
       eprint = {2510.20076},
 primaryClass = {astro-ph.HE},
       adsurl = {}
}

@ARTICLE{Soker2026G11,
       author = {{Soker}, Noam},
        title = "{Point-symmetric morphology in supernova remnant G11.2-0.3: the jittering jets explosion mechanism}",
      journal = {arXiv e-prints},
         year = 2026,
        month = nov,
          eid = {arXiv:2511.14578},
        pages = {arXiv:2511.14578},
          doi = {10.48550/arXiv.2511.14578},
archivePrefix = {arXiv},
       eprint = {2511.14578},
 primaryClass = {astro-ph.HE},
       adsurl = {https://ui.adsabs.harvard.edu/abs/2025arXiv251114578S}
}

@ARTICLE{Calvertetal2025,
       author = {{Calvert}, David and {Redle}, Michael and {Gautam}, Bibek and {Stapleford}, Charles J. and {Fr{\"o}hlich}, Carla and {Kneller}, James P. and {Liebendorfer}, Matthias},
        title = "{Turbulence in Core-Collapse Supernovae}",
      journal = {arXiv e-prints},
         year = 2025,
        month = nov,
          eid = {arXiv:2511.16755},
        pages = {arXiv:2511.16755},
archivePrefix = {arXiv},
       eprint = {2511.16755},
 primaryClass = {astro-ph.HE},
       adsurl = {https://ui.adsabs.harvard.edu/abs/2025arXiv251116755C}
}

@ARTICLE{SokerShiran2025,
       author = {{Soker}, Noam and {Shiran}, Kobi},
        title = "{Multiple shells in supernova 2023ixf support the jittering jets explosion mechanism (JJEM)}",
      journal = {arXiv e-prints},
         year = 2025,
        month = oct,
          eid = {arXiv:2510.18782},
        pages = {arXiv:2510.18782},
          doi = {10.48550/arXiv.2510.18782},
archivePrefix = {arXiv},
       eprint = {2510.18782},
 primaryClass = {astro-ph.HE},
       adsurl = {https://ui.adsabs.harvard.edu/abs/2025arXiv251018782S}
}

@ARTICLE{Dodson_etal_2003,
       author = {{Dodson}, R. and {Legge}, D. and {Reynolds}, J.~E. and {McCulloch}, P.~M.},
        title = "{The Vela Pulsar's Proper Motion and Parallax Derived from VLBI Observations}",
      journal = {\apj},
         year = 2003,
        month = oct,
       volume = {596},
       number = {2},
        pages = {1137-1141},
          doi = {10.1086/378089},
archivePrefix = {arXiv},
       eprint = {astro-ph/0302374},
 primaryClass = {astro-ph},
       adsurl = {https://ui.adsabs.harvard.edu/abs/2003ApJ...596.1137D}
}

@ARTICLE{Kochanek2022,
       author = {{Kochanek}, C.~S.},
        title = "{The progenitor of the Vela pulsar}",
      journal = {\mnras},
         year = 2022,
        month = apr,
       volume = {511},
       number = {3},
        pages = {3428-3439},
          doi = {10.1093/mnras/stac098},
archivePrefix = {arXiv},
       eprint = {2110.11369},
 primaryClass = {astro-ph.GA},
       adsurl = {https://ui.adsabs.harvard.edu/abs/2022MNRAS.511.3428K}
}

@ARTICLE{Ferrarietal2025,
       author = {{Ferrari}, L. and {Folatelli}, G. and {Ertini}, K. and {Kuncarayakti}, H. and {Regna}, T. and {Bersten}, M.~C. and {Ashall}, C. and {Baron}, E. and {Burns}, C.~R. and {Galbany}, L. and {Hoogendam}, W.~B. and {Maeda}, K. and {Medler}, K. and {Morrell}, N.~I. and {Shappee}, B. and {Stritzinger}, M.~D. and {Xiao}, H.},
        title = "{The nebular phase of SN 2024ggi: A low-mass progenitor with no signs of interaction}",
      journal = {\aap},
         year = 2025,
        month = oct,
       volume = {703},
          eid = {A12},
        pages = {A12},
          doi = {10.1051/0004-6361/202556652},
archivePrefix = {arXiv},
       eprint = {2507.22794},
 primaryClass = {astro-ph.SR},
       adsurl = {https://ui.adsabs.harvard.edu/abs/2025A&A...703A..12F}
}

@ARTICLE{Ertinietal2025,
       author = {{Ertini}, K. and {Regna}, T.~A. and {Ferrari}, L. and {Bersten}, M.~C. and {Folatelli}, G. and {Mendez Llorca}, A. and {Fern{\'a}ndez-Laj{\'u}s}, E. and {Ferrero}, G.~A. and {Hueichap{\'a}n D{\'\i}az}, E. and {Cartier}, R. and {Rom{\'a}n Aguilar}, L.~M. and {Putkuri}, C. and {Piccirilli}, M.~P. and {Cellone}, S.~A. and {Moreno}, J. and {Orellana}, M. and {Prieto}, J.~L. and {Gerlach}, M. and {Acosta}, V. and {Ritacco}, M.~J. and {Schujman}, J.~C. and {Vald{\'e}z}, J.},
        title = "{SN 2024ggi: Another year, another striking Type II supernova}",
      journal = {\aap},
         year = 2025,
        month = jul,
       volume = {699},
          eid = {A60},
        pages = {A60},
          doi = {10.1051/0004-6361/202554333},
archivePrefix = {arXiv},
       eprint = {2503.01577},
 primaryClass = {astro-ph.HE},
       adsurl = {https://ui.adsabs.harvard.edu/abs/2025A&A...699A..60E}
}

@ARTICLE{ChenTWetal2025,
       author = {{Chen}, Ting-Wan and {Yang}, Sheng and {Srivastav}, Shubham and {Moriya}, Takashi J. and {Smartt}, Stephen J. and {Rest}, Sofia and {Rest}, Armin and {Lin}, Hsing Wen and {Miao}, Hao-Yu and {Cheng}, Yu-Chi and {Aryan}, Amar and {Cheng}, Chia-Yu and {Fraser}, Morgan and {Huang}, Li-Ching and {Lee}, Meng-Han and {Lai}, Cheng-Han and {Liu}, Yu-Hsuan and {Sankar. K}, Aiswarya and {Smith}, Ken W. and {Stevance}, Heloise F. and {Wang}, Ze-Ning and {Anderson}, Joseph P. and {Angus}, Charlotte R. and {de Boer}, Thomas and {Chambers}, Kenneth and {Duan}, Hao-Yuan and {Erasmus}, Nicolas and {Fulton}, Michael and {Gao}, Hua and {Herman}, Joanna and {Hou}, Wei-Jie and {Hsiao}, Hsiang-Yao and {Huber}, Mark E. and {Lin}, Chien-Cheng and {Lin}, Hung-Chin and {Magnier}, Eugene A. and {Man}, Ka Kit and {Moore}, Thomas and {Ngeow}, Chow-Choong and {Nicholl}, Matt and {Ou}, Po-Sheng and {Pignata}, Giuliano and {Shiau}, Yu-Chien and {Sommer}, Julian Silvester and {Tonry}, John L. and {Wang}, Xiao-Feng and {Wainscoat}, Richard and {Young}, David R. and {Yeh}, You-Ting and {Zhang}, Jujia},
        title = "{Discovery and Extensive Follow-up of SN 2024ggi, a Nearby Type IIP Supernova in NGC 3621}",
      journal = {\apj},
         year = 2025,
        month = apr,
       volume = {983},
       number = {1},
          eid = {86},
        pages = {86},
          doi = {10.3847/1538-4357/adb428},
archivePrefix = {arXiv},
       eprint = {2406.09270},
 primaryClass = {astro-ph.HE},
       adsurl = {https://ui.adsabs.harvard.edu/abs/2025ApJ...983...86C}
}

@ARTICLE{Huetal2025,
       author = {{Hu}, Maokai and {Ao}, Yiping and {Yang}, Yi and {Hu}, Lei and {Li}, Fulin and {Wang}, Lifan and {Wang}, Xiaofeng},
        title = "{Early-time Millimeter Observations of the Nearby Type II SN 2024ggi}",
      journal = {\apjl},
         year = 2025,
        month = jan,
       volume = {978},
       number = {2},
          eid = {L27},
        pages = {L27},
          doi = {10.3847/2041-8213/ada1cd},
archivePrefix = {arXiv},
       eprint = {2412.11389},
 primaryClass = {astro-ph.SR},
       adsurl = {https://ui.adsabs.harvard.edu/abs/2025ApJ...978L..27H}
}
\bibliographystyle{aasjournal}
  


\end{document}